\documentclass[a4paper,11pt]{article}
\usepackage{pos} %
\usepackage[utf8]{inputenc} %
\usepackage{cleveref} %
\usepackage{siunitx} %
\DeclareSIUnit \parsec {pc}
\DeclareSIUnit \GeV {GeV}
\usepackage[margin=10pt]{subcaption} %

\usepackage{grffile} %

\usepackage{setspace}
\usepackage{wrapfig}
\usepackage{lineno}

\title{IceCube searches for GeV neutrino counterparts associated with high-energy starting events}
\ShortTitle{GeV neutrino counterparts to high-energy neutrinos}

\author{The IceCube Collaboration \\{\normalsize \normalfont(a complete list of authors can be found at the end of the proceedings)}\\}

\emailAdd{chraab@icecube.wisc.edu}
\emailAdd{gwenhael.dewasseige@uclouvain.be}

\abstract{The origin of the astrophysical neutrino flux observed by IceCube is largely unknown. To help decipher its astrophysical origin, we propose an IceCube analysis that conducts follow-up searches for GeV neutrinos associated with neutrino events above 60 TeV, which are known to have a high probability to be of astrophysical origin. 
It could not only identify a new component of the astrophysical neutrino flux, but also characterize how its spectrum extrapolates from GeV to PeV energies. This would in turn give valuable insights into the internal processes of neutrino sources. Astrophysical transients, such as collapsars, have been proposed as sources of time-correlated GeV- and high-energy neutrinos. Conducting this search in such short time scales allows for a substantial reduction in the dominant background rate for GeV neutrino candidate events. We introduce the statistical method and sensitivity of this search as well as dedicated data quality checks.
None of the searches yield statistically significant results, and we present the first limits on GeV neutrino emission associated to VHE neutrino events at short time scales.

\vspace{4mm}

{\bfseries Corresponding authors:}
Christoph Raab$^{1*}$,
Gwenhaël Wilberts de Wasseige$^{1}$\\
{$^{1}$ \itshape Centre for Cosmology, Particle Physics and Phenomenology -- CP3, Université catholique de Louvain}\\[4mm]
$^*$ Presenter
}

\ConferenceLogo{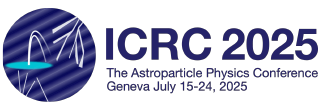}

\FullConference{39th International Cosmic Ray Conference (ICRC2025)\\
 15–24 July 2025\\
Geneva, Switzerland\\}

\begin{document}
\maketitle

\section{Motivation}
\label{sec:mot}

The IceCube neutrino observatory at the South Pole has observed a diffuse flux of astrophysical neutrinos at very high energies (above \qty{60}{TeV})~\cite{Lad:2025:AstrophysicalNeutrinoFlavor}, but so far all source classes that have been linked to neutrino observations are insufficient to explain the observed diffuse flux. 
Given the hypothesis that a single very-high-energy (VHE) neutrino could trace a transient neutrino production site, they are disseminated as public alerts to the multi-messenger community, to facilitate follow-up searches. These include looking for additional TeV--PeV neutrinos in IceCube's own data to identify potential neutrino transients~\cite{Meagher:2021:IceCubeMultimessengerFollowup,Pizzuto:2022:RealtimeFollowupAstrophysical}. %

We discuss in the following IceCube's potential to extend this principle down to GeV energies. Discovering counterparts in the GeV range would imply the presence of a new component of the astrophysical neutrino flux at GeV energies, and hence provide insights into the underlying physical processes. 

An absence of a detection conversely constrains the role of astrophysical models that predict GeV neutrinos in events that could also produce high-energy neutrinos. 
One example are choked-jet GRBs~\cite{Nakar:2015:UnifiedPictureLowluminosity,Senno:2016:ChokedJetsLowLuminosity}, where the GRB jet is stopped in extended material with a radius $r$ of $\SIrange{e13}{e14}{cm}\sim\SIrange{0.7}{6.7}{AU}$, whereby the electromagnetic emission evades detection and association with the neutrino event. These GRBs are also candidate cosmic ray accelerators. As the accelerated cosmic rays interact with matter or a target photon field, TeV neutrino emission is predicted from decay of the resulting pions.
The emission time scale $\Delta t$ can be estimated based on the size $r$ and Lorentz factor $\Gamma$, related by the condition $r = 2\Gamma^2 \Delta t$~\cite{Denton:2018:ExploringPropertiesChoked}, with a lower bound from the lifetime of the central engine. GeV neutrinos meanwhile are typically predicted from proton-neutron (p-n) collisions. These can for example be due to the collision of the neutron wind emitted by a collapsar, a rotating proto-neutron star (PNS), and occur typically on a time scale on the order of 10 seconds~\cite{Carpio:2024:QuasithermalGeVNeutrinos}. However, p-n collisions jet can also be related to decoupling within the jet~\cite{Murase:2022:NeutrinosBrightestGammaRay}.

Another case for combined low- and high-energy emission are short GRBs as a result of binary neutron star mergers. These events are also visible via gravitational waves prior to the emission of photons and neutrinos. In the case of of GRB~170818A/GW170817, IceCube has conducted %
GeV neutrino searches within 3 seconds of the merger~\cite{Kruiswijk:2023:Firstresultslowenergy}. If a population of such
GRBs exists where gamma rays are absorbed by the environment, GeV neutrino searches triggered by VHE neutrinos can still help reveal them. Unobscured GRBs meanwhile can not be considered candidates for this search as none have been associated to VHE neutrinos.

\section{Event selections at high and low energies}
\label{sec:dat}

The IceCube Observatory consists of a cubic kilometre of glacial ice instrumented with 5160 digital optical modules (DOMs) and a surface air shower array. It detects the Cherenkov light induced by relativistic charged particles, such as those produced by neutrino interactions, with nanosecond precision~\cite{IceCube:detectorpaper}. Most of the array has uniform horizontal and vertical spacing, while a more densely spaced array, called IceCube-DeepCore~\cite{IceCube:detectorpaper}, is located in the deepest and clearest region of the ice and provides sensitivity to GeV neutrinos.

In the following we describe the two neutrino event selections used in the analysis presented in these proceedings.

\paragraph{The High-Energy Starting Event (HESE) sample} contains events produced by some of the most energetic neutrino interactions detected by IceCube. It achieves this with simple cuts on the total light yield while rejecting atmospheric muons that produce light in the outer regions of the detector. Furthermore requiring a reconstructed deposited energy above \SI{60}{TeV} results in $\sim84\%$ of its events originating from astrophysical sources~\cite{Lad:2025:AstrophysicalNeutrinoFlavor}.
In the following, we will use 97 events observed in the HESE sample with 12 years of IceCube data as a sample of VHE events to search for coincident low energy neutrinos. %
As in previous spatial clustering analyses, two HESE events identified as contaminated by atmospheric background were removed from the sample~\cite{IceCube:2024:hese12}.

\paragraph{IceCube's Extremely LOW ENergy (ELOWEN) selection} contains the lowest energy events able to trigger the detector, with individual photons detected in a small number of DOMs. It makes use of spatial and temporal causality conditions between a small number of modules in IceCube-DeepCore to identify the light deposited by the charged particles propagating from a GeV neutrino interaction. This enables distinguish such a signal from detector noise which is uncorrelated between DOMs. The dominant background for IceCube events are higher-energy atmospheric muons. These are rejected via the charge deposited in the surrounding array. The selection also limits the number of DOMs detecting a possible photon within IceCube-DeepCore. ELOWEN is dominated by events purely due to detector noise. Its rate of \SI{20}{mHz}~\cite{Abbasi:2021:SearchGeVNeutrino} allows for analyses focused on short transients~\cite{IceCubeCollaboration:2023:LimitsNeutrinoEmission,Kruiswijk:2023:Firstresultslowenergy}. Assuming an $E_\nu^{-2}$ spectrum, it accepts neutrinos with energies from \SIrange{2.6}{64}{GeV} (central 90\%). %
Presently, ELOWEN events are not reconstructed in terms of neutrino arrival direction. The following analysis therefore only makes use of the temporal correlation, assuming transient sources.

\paragraph{The quality of IceCube data} is validated with extensive monitoring~\cite{IceCube:detectorpaper} applied to data-taking periods typically lasting eight hours. If necessary, a data-taking period can either be marked as bad, or only an interval of it declared as suitable for analysis.
The analysis shown here includes only good data-taking periods in its livetime. Furthermore, following the procedure in the follow-up of gravitational wave events with ELOWEN~\cite{Kruiswijk:2023:Firstresultslowenergy}, we require operation of the complete detector configuration. For a small number of data-taking periods before 2016, the available data is incomplete. In order to avoid dealing with such gaps, these are also excluded.

\paragraph{Each event's proposed follow-up must fulfil certain criteria} in order to proceed. Foremost, the entire data-taking period containing the HESE event must pass the data-taking period selection described in \Cref{sec:dat}, which excludes three analyses from the period when IceCube consisted of 79 strings, and two from data-taking periods whose data is not complete.
The analyses described in \Cref{sec:cou} and \Cref{sec:tim} use an off-time window preceding the HESE event to obtain a background rate estimate. We ensure that a minimum 4 hours of livetime are available for this after the data-taking period selection, and consistent rates of \SIrange{17}{22}{mHz} following the condition used in Reference~\cite{Kruiswijk:2023:Firstresultslowenergy}. This precludes the follow-up of 14 HESE events due to elevated background rates. 
We require smooth data-taking period transitions within the time windows where searches for GeV neutrinos are performed, as discussed in \Cref{sec:res}.

\section{Counting analyses}
\label{sec:cou}

We conduct a model-independent analysis to identify excesses of low-energy neutrino data within a fixed time window around a HESE event. We use the same statistical analysis method and framework as the ELOWEN O4 follow-up~\cite{Kruiswijk:2023:Firstresultslowenergy}. %
This method employs an off-time window of 8 hours preceding the transient to provide a background rate estimate for the analysis period. This is then used to calculate the significance of the observation (according to Section 3 of Reference~\cite{Cousins:2008:EvaluationThreeMethods}). It also derives a Bayesian upper limit from combining the off-time measurement with an unbiased prior on the rate~\cite{Kruiswijk:2023:Firstresultslowenergy}. %
When none of the analyses are individually significant, a signal present in multiple follow-ups could still cause a significant excess when considered as a population. This is examined with a binomial test, already employed in the ELOWEN gravitational wave follow-up~\cite{Kruiswijk:2023:Firstresultslowenergy}. %

The main time window of the analysis is is \SIrange{-500}{500}{s}. This broad window is designed to cover diverse emission scenarios. For instance in the simple model outlined in \Cref{sec:mot}, such TeV emission time scales could be achieved when the jet has a low Lorentz factor $\Gamma\lesssim10$, not accounting for the active time of the central engine.  
It also encompasses the time scale encouraged for a search for GeV--TeV neutrinos from GRB~221009A~\cite{Murase:2022:NeutrinosBrightestGammaRay}.
In addition, we search for GeV neutrinos within a window of \SIrange{-3}{3}{s} to improve the sensitivity towards transients on this even shorter scale%
due to reduced background, as shown in \Cref{fig:summary}.

\section{Timing analysis}
\label{sec:tim}

When an astrophysical transient at time $t_0$ produces GeV neutrinos, the times $t$ of resulting ELOWEN events will be distributed according to a probability density function (PDF) $P_\text{GeV}(t - t_0)$ with a characteristic internal time scale $\Delta t$. 
Since $t_0$ is unknown, we instead analyse the time differences $(t - t_\text{HESE})$. Their distribution is then characterised by the internal scale $\Delta t$, at which possible ELOWEN multiplets would cluster; and the VHE emission time scale $\Delta t_\text{HESE}$, which determines the position in the time window.

A timing analysis of ELOWEN events within the main 1000-second time window discussed in \Cref{sec:cou} can provide improved sensitivity towards faster variability of the GeV emission compared to a simple counting analysis, and should be robust against when it arrives in this window.

\paragraph{Our timing analysis method} is based on one specific to short transients called PeaNuTS~\cite{Lamoureux:2023lgq}. 
It examines the time differences $t_{i+1} -t_i$ between subsequent events in a pre-defined time window. Assuming that the events follow a constant background rate $r_\text{bkg}$, their time differences are expected to follow an exponential distribution. %
The present work assigns each event $i$ a test statistic according to the immediately preceding $i-1$ and subsequent $i+1$ event:
\begin{equation}
\text{TS}_i = - \log\left[{1 - e^{-r_\text{bkg}(t_i - t_{i-1})}}\right] - \log\left[{1 - e^{-r_\text{bkg}(t_{i+1} - t_{i})}}\right] .
\end{equation}

\begin{figure}
\begin{minipage}[t]{0.48\linewidth}    
     \centering
    \includegraphics[width=\columnwidth]{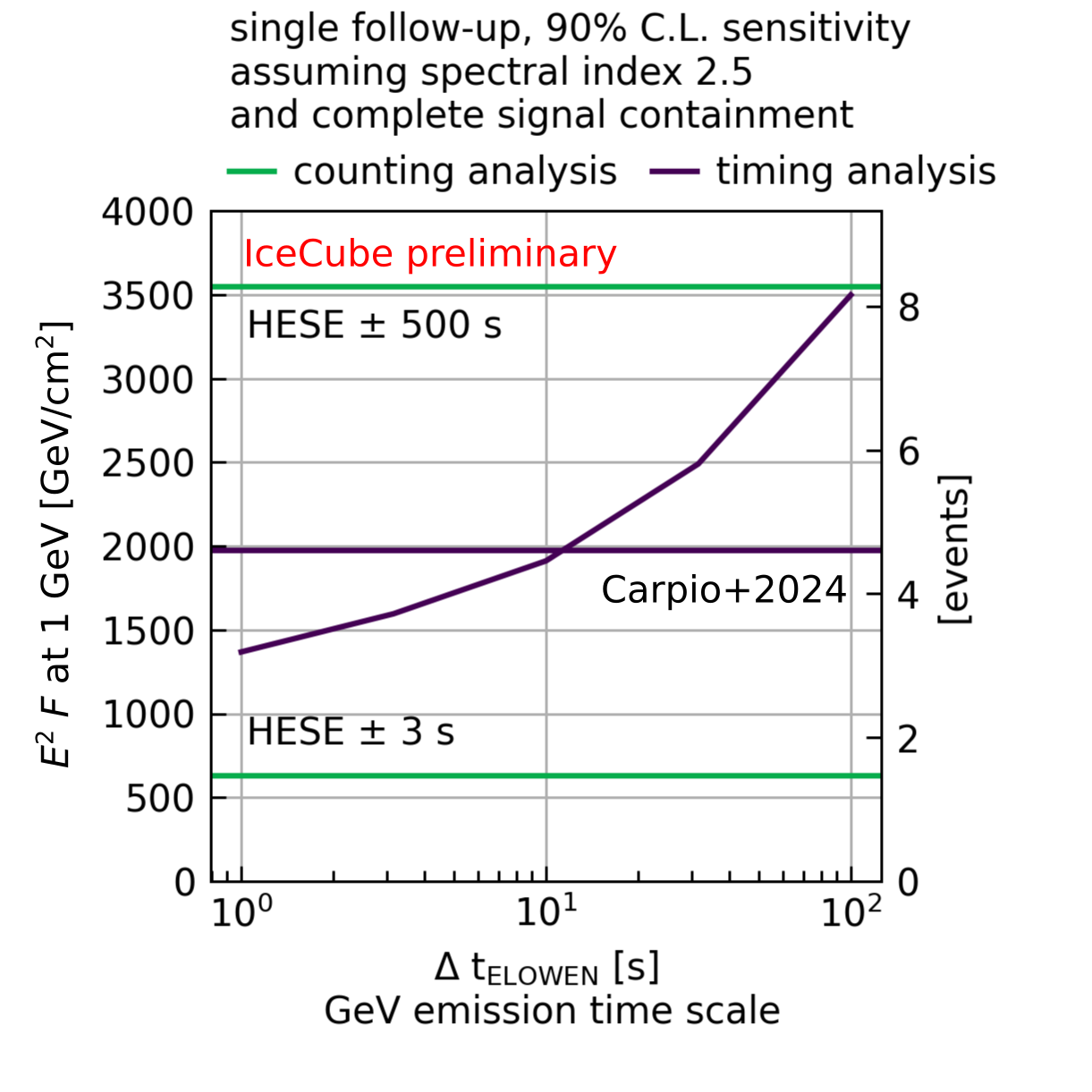}
    \caption{Sensitivity at 90\% C.L. of the different analyses performed. For the counting analysis, 100\% signal containment in the respective time window is assumed. For the timing analysis, the GeV emission follows either an exponential template with a variable time scale (x-axis) or a physical model from Carpio et al.~\cite{Carpio:2024:QuasithermalGeVNeutrinos}. In the left-hand y-axis, F refers to time-integrated all-flavour neutrino + anti-neutrino flux. }
    \label{fig:summary}
    \end{minipage}
\hfill
\begin{minipage}[t]{0.48\linewidth}    
     \centering
    \includegraphics[width=1.05\columnwidth]{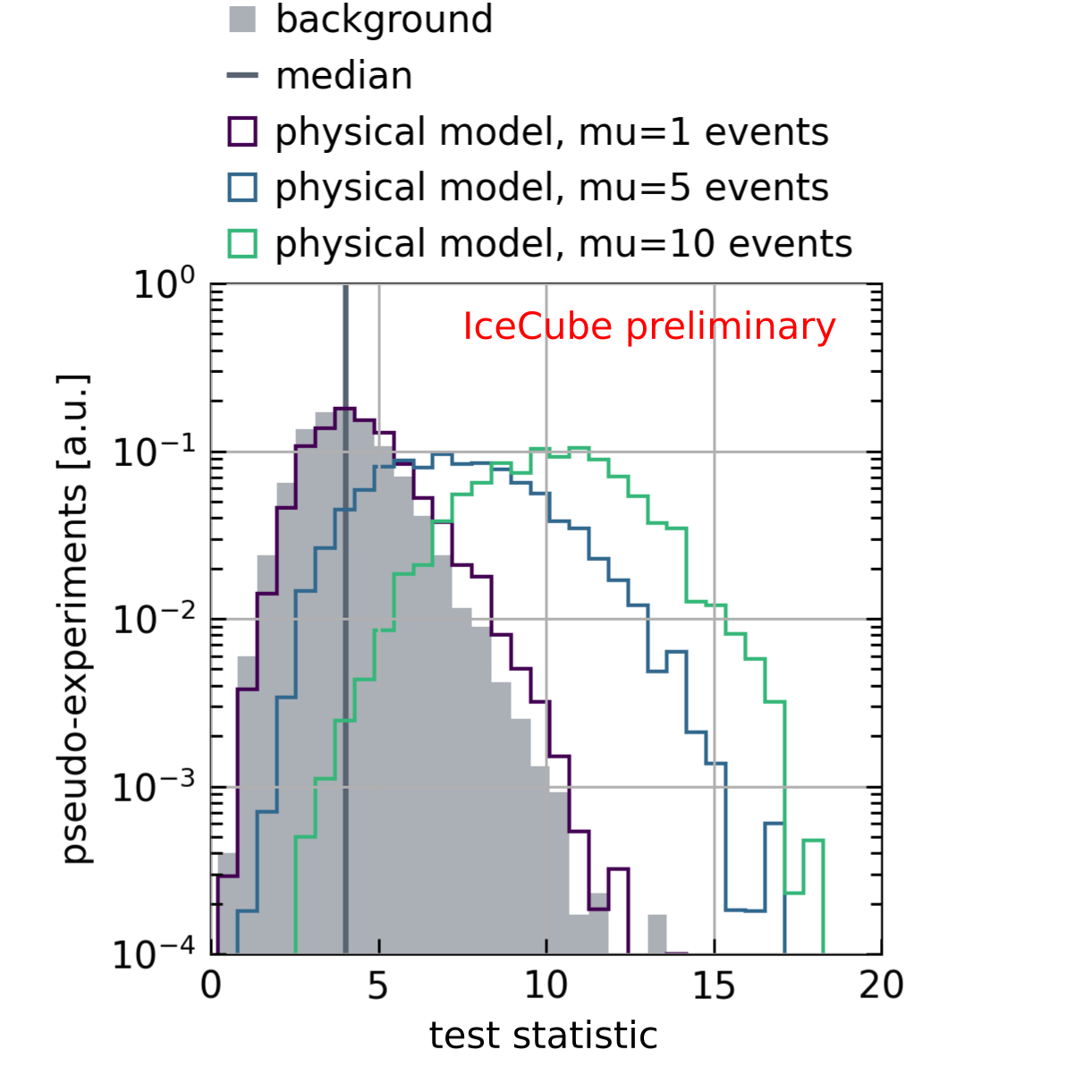}
    \caption{Distribution of the timing analysis test statistic under toy Monte Carlo for the background hypothesis (grey) and injecting according to a physical model at three signal strengths.}
    \label{fig:tim/demo}
    \end{minipage}
\end{figure}

This event-wise test statistic increases both when neutrino events cluster at shorter time scales and in larger multiplicity. The original PeANuTS performs a hypothesis test by defining a threshold value of $TS_{3\sigma}$ for one event and then requiring that two events $\text{TS}_i\geq\text{TS}_{3\sigma}$ in order to suppress the effect of a chance background event near one due to signal. In analogy to this approach, we define the timing analysis test statistic as the second-largest element of the set $\lbrace\text{TS}_i\rbrace$. 
The effect of injecting a signal on the distribution is shown in \Cref{fig:tim/demo}.

\paragraph{The performance of this method} is presented in terms of the median sensitivity at 90\% confidence level (CL) towards both a generic and a physical hypothesis. We employ the Neyman construction on ensembles of pseudo-experiments generated with a toy Monte Carlo representing these signal hypotheses together with a steady background of \SI{20}{mHz}. %

For the generic hypothesis, we assume that both the GeV and VHE neutrino light curves begin at the same unknown transient time, after which they decay according to a natural exponential $\propto \exp{-\frac{t - t_0}{\Delta t}}$ with respective decay constants  of $\Delta t$ and $\Delta t_\text{HESE}$.  We study the sensitivity depending on $\Delta t$ in a range of \SIrange{1}{100}{s} and show the results in the purple curves in \Cref{fig:summary}. At the lower end, the method approaches its maximum sensitivity, while at the upper range, it falls behind the counting analysis. 

For an example of a physical model, we consider the GeV neutrino light curves predicted by Reference~\cite{Carpio:2024:QuasithermalGeVNeutrinos}, all of which decay approximately exponentially. Their decay time varies by only 20\% around 10 seconds. We choose the model corresponding to a proto-neutron star with a surface magnetic field of \SI{e15}{G} and a rotation period of \SI{1}{s} as both representative of the range of decay scales, and predicting the largest neutrino yield in the GeV band. As the estimated GeV decay time lies within the range studied for the generic hypothesis, so does the sensitivity shown in \Cref{fig:summary}.

In both hypotheses, we compare a large delay $\Delta t_\text{HESE}=\SI{100}{s}$ with prompt emission ($\Delta t_\text{HESE}=\SI{0.1}{s}$). This serves to assure ourselves the method is as robust as intended with respect to the delay between the HESE event and ELOWEN cluster.

\section{Post-unblinding check}
\label{sec:post}

It is possible that individual strings or DOMs exhibit an elevated noise rate which is too short in duration to be detected by the data-taking period-wise monitoring, but coincides with the signal window. This is relevant to the analysis in case these same additional noise hits trigger data acquisition, causing events which survive at the final event selection. This is not equally likely for all DOMs, owing to the explicit definition of the trigger condition and the event selection using outer regions of the array to reject entering atmospheric muons, leaving inner fiducial regions to be examined for potential neutrino interactions. We therefore perform a post-unblinding check which counts how often each DOM $i$ and string $j$ is represented among the DOMs
that
are likely to have caused the triggers to take data during the analysis window.

\begin{figure}
\begin{minipage}[t]{0.48\linewidth}    
    \includegraphics[width=\linewidth]{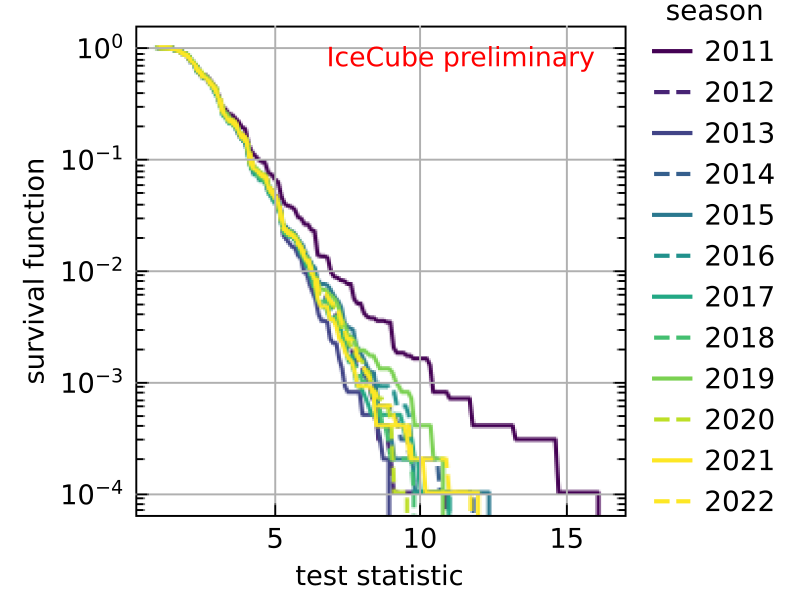}
    \caption{Survival function of the string-wise noise excess TS distribution obtained from sampling random sets of 20 events from a particular season (colour scale) and performing the test described in \Cref{sec:post}. The 2011 data-taking season follows the deployment of the final IceCube strings, and the decreasing noise rate is thought to be an indirect effect of refreezing~\cite{IceCube:2024:SearchGalacticCorecollapse}.}
    \label{fig:dat/post}
\end{minipage}
\hfill
\begin{minipage}[t]{0.48\linewidth}
    \centering
    \includegraphics[width=0.9\linewidth]{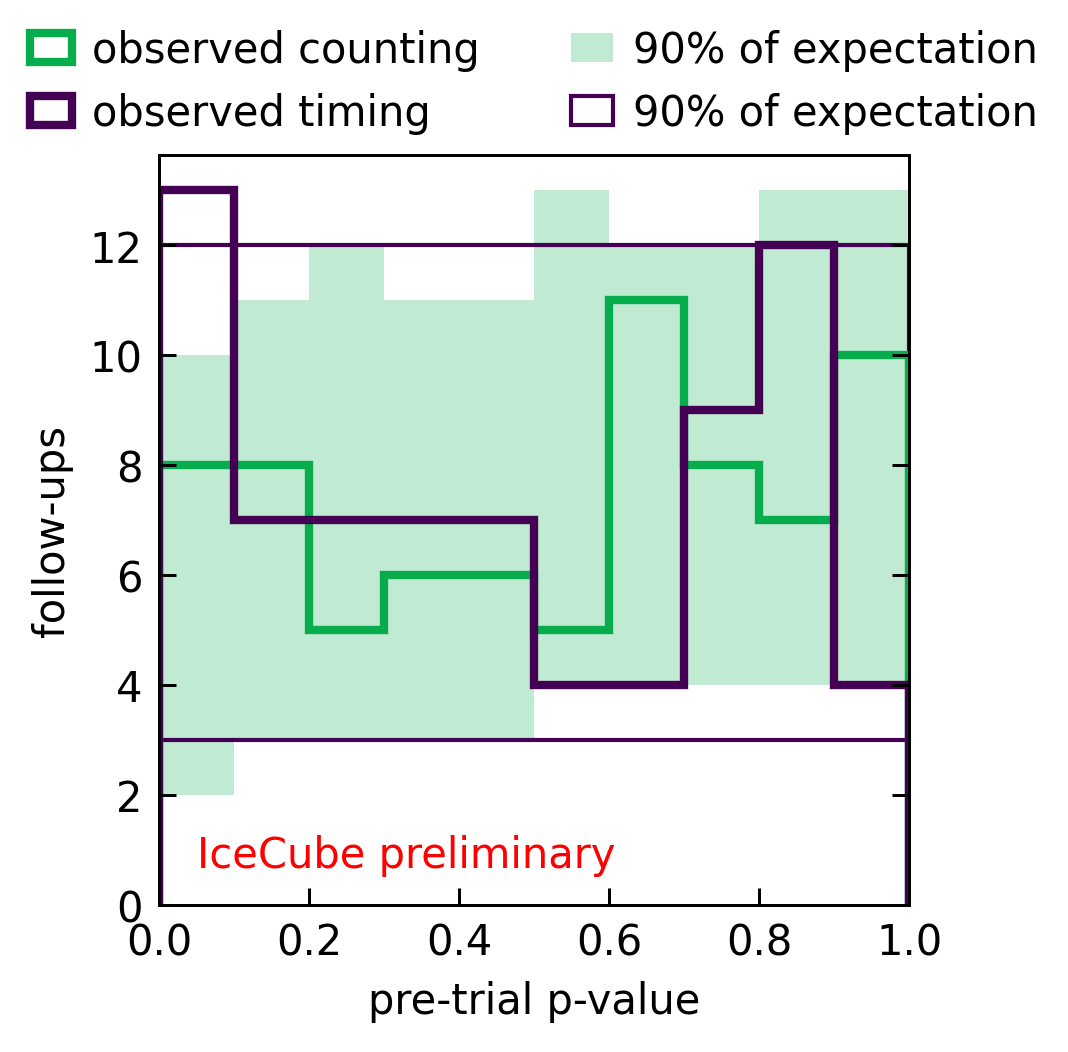}
    \caption{Histogram of p-values resulting from the 74 counting (green) and timing (purple) analyses with a 1000-second time window. The expected range of each histogram bin (number of follow-ups) is indicated by the respective shaded and outlined regions, if each follow-up follows its background expectation.}
    \label{fig:res/p}
    \end{minipage}
\end{figure}

We build background histograms from ELOWEN data during the first of each month of each year. The counts per DOM $N_i$ or string $N_j$ are normalised by the number of ELOWEN events $N_\text{background}$ to $p_i=N_i/N_\text{background}$ and $p_j=N_j/N_\text{background}$. 
These are then compared to the counts $N_i$ and  $N_j$ among all $N_\text{obs}$ events in the unblinded time window. An excess is identified via the one-sided Poissonian significance $P(N_i \geq p_i N_\text{obs})$ or $P(N_j \geq p_j N_\text{obs})$. The test statistic is $-\log_{10} \hat{P}$ with $\hat{P}$ the most significant such excess. Pseudo experiments for the null hypothesis are generated by performing the check on a set of $N_\text{obs}$ events, randomly selected from the background sample matching the same data-taking season. 
These pseudo-experiments define the final significance reported by the checks, and their distributions for $N_\text{obs}=20$ are shown in \Cref{fig:dat/post}.

\section{Results and conclusions}
\label{sec:res}

Four of the 1000-second time windows overlap with the stop of one data-taking period and start of the next. While for two this is only a logical transition, the other two include an interruption in the data taking of respectively 86 and 91 seconds. %
To avoid possibly different noise behaviour at the beginning of a full start, the long time window is not analysed, which also helps achieve more consistency in the timing analysis. However, the short time windows conclude before the end of the respective data-taking period and therefore their counting analysis does proceed.

In total, the counting and timing analysis on the 1000-second time window was unblinded for 74 HESE events. These p-values are histogrammed in \Cref{fig:res/p}, where they are compatible with the statistical fluctuations from a Poissonian background for the same number of follow-ups. 
The counting p-values for the 74 1000-second time windows range between 1.4\% and 97.8\% and are therefore consistent with background. The limits on the $E^2 F$ at \SI{1}{GeV}, with $F$ the time-integrated flux of a spectrum proportional to $E^{-2.5}$, range from $\SIrange{1886}{8307}{GeV/cm^2}$. The 6-second counting analysis always finds 0 or 1 events, which is also consistent with background and the corresponding limits %
are either around $\SI{622}{GeV/cm^2}$ or $\SI{1275}{GeV/cm^2}$, depending slightly on the given background level. These are the first limits on GeV neutrino emission associated with astrophysical neutrinos.
The binomial test finds a post-trial p-value of 33.2\% from the 16 most significant 1000-second counting analyses. 

The timing analysis (see \Cref{sec:tim}) was applied to the same events which were followed up with the 1000-second counting, within the same window. The most significant of the p-values trial-corrects to $p_\text{post} = 1 -\left(1-0.0057\right)^{74} = 34.6\%$\footnote{the \v{S}id\'{a}k correction}. For the most significant test, the multiplets are located at \SI{302.57}{s} and \SI{302.44}{s} after the HESE events.

The post-unblinding check described in \Cref{sec:post} yields significance values $\geq1.8\%$ overall. It is less significant still for the follow-ups contributing to the binomial test and post-trial timing p-value. We therefore find no indication of a noise excess driving these results in the first study of GeV neutrino counterparts to a highly astrophysical sample of neutrinos observed by IceCube. No such component has been identified, and the limits on neutrino emission at time scales below 1000~seconds are under $\SI{8307}{GeV/cm^2}$ for an $E^{-2.5}$ spectrum. The role of the candidate transients outlined in \Cref{sec:mot} for astrophysical neutrino production remains unknown. 

Future studies can be undertaken to examine a larger sample of candidate astrophysical neutrino events and their GeV follow-up in population analysis, taking into account the event parameters and likelihood to arise from the astrophysical flux. This will also to translate the follow-ups into limits on a population of GeV-VHE transients in a physically motivated way. Finally, future iterations of this analysis will benefit from improved sensitivity at GeV energies thanks to improved event reconstruction, event selection, and the installation of the IceCube Upgrade.

\bibliographystyle{JHEP}
\bibliography{elowen-hese}

\clearpage

\section*{Full Author List: IceCube Collaboration}

\scriptsize
\noindent
R. Abbasi$^{16}$,
M. Ackermann$^{63}$,
J. Adams$^{17}$,
S. K. Agarwalla$^{39,\: {\rm a}}$,
J. A. Aguilar$^{10}$,
M. Ahlers$^{21}$,
J.M. Alameddine$^{22}$,
S. Ali$^{35}$,
N. M. Amin$^{43}$,
K. Andeen$^{41}$
% ...
\\
\\
% ...
$^{10}$ Universit{\'e} Libre de Bruxelles, Science Faculty CP230, B-1050 Brussels, Belgium \\
% ...
$^{16}$ Department of Physics, Loyola University Chicago, Chicago, IL 60660, USA \\
$^{17}$ Dept. of Physics and Astronomy, University of Canterbury, Private Bag 4800, Christchurch, New Zealand \\
% ...
$^{21}$ Niels Bohr Institute, University of Copenhagen, DK-2100 Copenhagen, Denmark \\
$^{22}$ Dept. of Physics, TU Dortmund University, D-44221 Dortmund, Germany \\
% ...
$^{35}$ Dept. of Physics and Astronomy, University of Kansas, Lawrence, KS 66045, USA \\
% ...
$^{39}$ Dept. of Physics and Wisconsin IceCube Particle Astrophysics Center, University of Wisconsin{\textemdash}Madison, Madison, WI 53706, USA \\
% ...
$^{41}$ Department of Physics, Marquette University, Milwaukee, WI 53201, USA \\
% ...
$^{43}$ Bartol Research Institute and Dept. of Physics and Astronomy, University of Delaware, Newark, DE 19716, USA \\
% ...
$^{63}$ Deutsches Elektronen-Synchrotron DESY, Platanenallee 6, D-15738 Zeuthen, Germany \\
$^{\rm a}$ also at Institute of Physics, Sachivalaya Marg, Sainik School Post, Bhubaneswar 751005, India \\
% ...

\subsection*{Acknowledgments}

\noindent
The authors gratefully acknowledge the support from the following agencies and institutions:
USA {\textendash} U.S. National Science Foundation-Office of Polar Programs,
U.S. National Science Foundation-Physics Division,
U.S. National Science Foundation-EPSCoR,
U.S. National Science Foundation-Office of Advanced Cyberinfrastructure,
Wisconsin Alumni Research Foundation,
Center for High Throughput Computing (CHTC) at the University of Wisconsin{\textendash}Madison,
Open Science Grid (OSG),
Partnership to Advance Throughput Computing (PATh),
Advanced Cyberinfrastructure Coordination Ecosystem: Services {\&} Support (ACCESS),
Frontera and Ranch computing project at the Texas Advanced Computing Center,
U.S. Department of Energy-National Energy Research Scientific Computing Center,
Particle astrophysics research computing center at the University of Maryland,
Institute for Cyber-Enabled Research at Michigan State University,
Astroparticle physics computational facility at Marquette University,
NVIDIA Corporation,
and Google Cloud Platform;
Belgium {\textendash} Funds for Scientific Research (FRS-FNRS and FWO),
FWO Odysseus and Big Science programmes,
and Belgian Federal Science Policy Office (Belspo);
Germany {\textendash} Bundesministerium f{\"u}r Forschung, Technologie und Raumfahrt (BMFTR),
Deutsche Forschungsgemeinschaft (DFG),
Helmholtz Alliance for Astroparticle Physics (HAP),
Initiative and Networking Fund of the Helmholtz Association,
Deutsches Elektronen Synchrotron (DESY),
and High Performance Computing cluster of the RWTH Aachen;
Sweden {\textendash} Swedish Research Council,
Swedish Polar Research Secretariat,
Swedish National Infrastructure for Computing (SNIC),
and Knut and Alice Wallenberg Foundation;
European Union {\textendash} EGI Advanced Computing for research;
Australia {\textendash} Australian Research Council;
Canada {\textendash} Natural Sciences and Engineering Research Council of Canada,
Calcul Qu{\'e}bec, Compute Ontario, Canada Foundation for Innovation, WestGrid, and Digital Research Alliance of Canada;
Denmark {\textendash} Villum Fonden, Carlsberg Foundation, and European Commission;
New Zealand {\textendash} Marsden Fund;
Japan {\textendash} Japan Society for Promotion of Science (JSPS)
and Institute for Global Prominent Research (IGPR) of Chiba University;
Korea {\textendash} National Research Foundation of Korea (NRF);
Switzerland {\textendash} Swiss National Science Foundation (SNSF).

\end{document}